\documentclass{aa}  

\usepackage{graphicx}
\usepackage{txfonts}
\newcommand{\teff}{$T_{\rm eff}$}

\newcommand{\msun}{$M_\odot$}

\def\vsini{$v\sin i$}
\def\prot{${\rm P}_{rot}$}

\def\ha10{${\rm H}\alpha\,10\%$}

\begin{document}

   \title{The lithium-rotation connection in the 125 Myr-old Pleiades cluster\thanks{Based on observations made at Observatoire de Haute Provence (CNRS), France, at the Nordic Optical Telescope (IAC), Spain, and at the W. M. Keck Observatory, Hawaii, USA.}}


   \author{J. Bouvier
          \inst{1,2}
          \and
          D. Barrado\inst{3}
          \and 
        E. Moraux \inst{1}
           \and
          J. Stauffer \inst{4}
          \and
          L. Rebull \inst{4}
                  \and 
          L. Hillenbrand \inst{5}
        \and
          A. Bayo \inst{2}
          \and
          I. Boisse \inst{6} 
          \and
          H. Bouy \inst{7} 
          \and
          E. DiFolco \inst{7}
          \and
          J. Lillo-Box \inst{8}
          \and
          M. Morales Calder\'{o}n \inst{3}
        }

 \institute{Univ. Grenoble Alpes, CNRS, IPAG, F-38000 Grenoble, France \\     \email{Jerome.Bouvier@univ-grenoble-alpes.fr}
     \and
              Instituto de F\'{\i}sica y Astronom\'{\i}a, Universidad de Valpara\'{\i}so, Chile
              \and
              Depto. Astrof\'{\i}sica, Centro de Astrobiolog\'{\i}a (INTA-CSIC), Spain
              \and   
              Spitzer Science Center, California Institute of Technology, Pasadena, CA 91125, USA
              \and
                        Department of Astronomy, California Institute of Technology, Pasadena, CA 91125, USA
                        \and
                  Aix Marseille Universit\'e, CNRS, Laboratoire d'Astrophysique de Marseille UMR 7326, 13388 Marseille, France
                  \and
           Laboratoire d'Astrophysique de Bordeaux, Univ. Bordeaux, CNRS, 33615 Pessac, France
           \and European Southern Observatory, Alonso de Cordova 3107, Vitacura Casilla 19001, Santiago 19, Chile
            }

   \date{Received ; accepted}

 
  \abstract
      {The evolution of lithium abundance over a star's lifetime is indicative of transport processes operating in the stellar interior.}
      {We revisit the relationship between lithium content and rotation rate previously reported for cool dwarfs in the Pleiades cluster. }
   {We derive new LiI 670.8 nm equivalent width measurements from high-resolution spectra obtained for low-mass Pleiades members. We combine these new measurements with previously published ones, and use the Kepler/K2 rotational periods recently derived for Pleiades cool dwarfs to investigate the lithium-rotation connection in this 125 Myr-old cluster.}
   { The new data confirm the correlation between lithium equivalent width and stellar spin rate for a sample of 51 early K-type members of the cluster, where fast rotating stars are systematically lithium-rich compared to slowly rotating ones. The correlation is valid for {\it all} stars over the (J-K$_s$) color range 0.50-0.70 mag, corresponding to a mass range from about 0.75 to 0.90 \msun, and may extend down to lower masses.
    }
   {We argue that the dispersion in lithium equivalent widths observed for cool dwarfs in the Pleiades cluster reflects an intrinsic scatter in lithium abundances, and suggest that the physical origin of the lithium dispersion pattern is to be found in the pre-main sequence rotational history of solar-type stars.  }


   \maketitle
%

\section{Introduction}

Stellar physics offers many puzzling issues to astronomers, and the evolution of lithium abundances in solar-type stars is probably one of the most demanding. A fragile element that burns at a temperature of 2.5 MK encountered at the base of the convective envelope of low-mass stars, lithium is slowly depleted during the star's residence on the main sequence \citep[see, e.g.,][for a recent review]{Jeffries14}. Beyond this secular trend, the depletion rate is extremely sensitive to temperature \citep[e.g.,][]{Bildsten97} and lithium abundances therefore provide an excellent probe to transport processes at work in stellar interiors: the more vigorous the transport of chemical elements to the hot central regions of the star, the more dramatic the lithium depletion. Indeed, various physical processes impact directly on the rate of internal transport, such as rotation or magnetic field. As an illustration, the recent asteroseismic measurements of the core rotation of stars on the red giant branch \citep[e.g.,][]{Deheuvels14} clearly highlight the discrepancy between actual rotational profiles and current model predictions that underestimate the efficiency of angular momentum transport by orders of magnitude \citep[e.g.,][]{Eggenberger12a, Marques13}. 

Indirect diagnostics of internal processes are therefore extremely valuable at all phases of stellar evolution, and surface lithium abundance is one of the most accessible through the LiI 670.8 nm photospheric feature present in the optical spectra of cool dwarfs. Over the past decades, this has motivated thorough investigations of the connection between lithium and rotation and/or magnetic activity in solar-type stars at various stages of evolution. Early studies of low-mass stars in young open clusters, for example, Pleiades \citep{Butler87} and Alpha Per \citep{Balachandran88}, surprisingly reported that slowly rotating late-type dwarfs reaching the zero-age main sequence were systematically lithium depleted compared to fast rotators of the same spectral type. \citet{Soderblom93} performed a quantitative study based on a large sample of cool dwarfs in the Pleiades cluster and clearly confirmed the earlier claims of a lithium spread among the low mass members. Specifically, while the G dwarf members of the cluster exhibit little spread in their lithium abundances, usually being close to the solar meteoritic abundance, they showed that K dwarfs on the contrary display a huge lithium spread, reaching up to 1 dex in abundance and increasing towards lower masses. Thus, they showed that among cluster members less massive than about 0.9\msun, those with the highest projected rotational velocities were systematically lithium-rich compared to their similar-mass, low-velocity siblings.  Subsequent studies confirmed these early results and showed that the lithium spread affects mostly early- to mid-K stars in the Pleiades cluster \citep[e.g.,][]{Russell96, King00, Barrado16}, while late-K and M dwarfs are heavily lithium depleted \citep[e.g.,][]{Garcialopez94, Oppenheimer97} down to near the hydrogen burning mass limit where lithium starts to reappear \citep{Basri96, Rebolo96, Stauffer98}.  Since then, additional examples of a lithium-rotation connection among low-mass dwarfs have been reported for young open clusters \citep[e.g., M34 and NGC 2516,][respectively]{Jones97, Jeffries98}, young moving groups \citep[e.g., $\beta$ Pic,][]{Messina16}, and even for star forming regions as young as 5 Myr \citep[e.g., NGC 2264,][]{Bouvier16}. 
  
The reason why fast rotators tend to be lithium-rich compared to slow ones in young clusters has prompted various interpretations. Rotation may directly impact the lithium depletion rate in stars with thick convective envelopes, either through structural effects during the pre-main sequence \citep[e.g.,][]{Martin96} or by reducing the penetration of convective plumes into the radiative core, which yields a lower depletion rate in fast rotators \citep{Siess97, Montalban00,  Baraffe17}. Alternatively, pre-main sequence angular momentum loss due to disk locking can build up a lithium dispersion pattern early on, by enhancing lithium depletion in slowly rotating stars with long-lived disks \citep{Bouvier08, Eggenberger12b}. Non-steady disk accretion during the early phases of stellar evolution can also trigger lithium over-depletion \citep{Baraffe10}, although the link with rotation then becomes more indirect. Finally, stellar magnetic fields may promote high lithium content by reducing the rate of energy transport in the outer convective envelope \citep{Ventura98, Feiden16}, leading to an inflated stellar radius and a lower temperature at the base of the convective zone \citep{Somers14, Somers15, Somers17, Jeffries17}. 


With the aim to further document the relationship between lithium content and rotation rate in young low-mass stars, we report here new lithium measurements for low-mass members of the Pleiades cluster and use the most recent and robust determinations of their rotational periods to revisit the lithium-rotation connection in this benchmark cluster. In Section 2, we introduce the stellar sample under consideration. In Section 3, we describe the spectroscopic observations we performed  to secure new measurements of lithium content for early K-type cluster members. In Section 4, using the rotational period measurements recently derived from Kepler/K2 studies of the cluster, we report the results regarding the link between rotation rate and lithium content among the FGK stars of the cluster, and delimit its domain of validity. In Section 5, we discuss the implications of the results for the internal transport processes. We conclude in Section 6 that there is a need to improve the characterization of the lithium-rotation connection in pre-main sequence clusters at various ages. 
   


\section{The sample}

%

\begin{table*}
\caption{Stellar parameters and new measurements of lithium equivalent width and abundance. Names are listed in alphabetical order.}             
\label{linew}      
\centering                          
\begin{tabular}{l l l l l l l l l}        
\hline\hline                 
Name & K2 EPIC\# & J-K$_s$ & P$_{rot}^\dagger$ & EW(LiI) & $\sigma_{EW(LiI)}$ & \teff & A[Li] & $\sigma_{A[Li]}$\\
     &  & {\it (mag)}  & \it{(d)} & \it{(m\AA)} & \it{(m\AA)} &\it{(K)} \\
\hline                        
   --      & 210942842 & 0.637 & 6.2883 & 187.0 & 10.0   & 4740    & 1.995  & 0.093  \\   
AKIV-314 & 211085419 & 0.659 & 8.064516 & 12.0 & 10.0 & 4684 & 0.363 & 0.163\\
DH 156 & 211039307 & 0.65 & 8.333334 & $<$15.0 &  -- & 4708 & 0.498 & 0.092\\
DH 343 & 210754915 & 0.582 & 0.533049 & 231.0 & 15.0 & 4970 & 2.567 & 0.095\\
DH 800 & 211153286 & 0.675 & 7.692308 & 124.0 & 13.0 & 4642 & 1.501 & 0.096\\
HII 659 & 211028956 & 0.644 & 2.808989 & 197.0 & 26.0 & 4723 & 2.026 & 0.106\\
PELS 019 & 211095259 & 0.524 & 6.849315 & 49.0 & 5.0 & 5168 & 1.622 & 0.079\\
PELS 030 & 210995339 & 0.567 & 6.666667 & 148.0 & 8.0 & 5011 & 2.122 & 0.085\\
PELS 031 & 210966700 & 0.613 & 5.0 & 221.0 & 11.0 & 4851 & 2.342 & 0.091\\
PELS 066 & 210922592 & 0.635 & 7.575758 & 60.0 & 9.0 & 4747 & 1.207 & 0.091\\
PELS 069  & 210924735 & 0.642 & 1.7525 & 217.0 & 10.0   & 4728   &  2.153  & 0.094\\  
PELS 071 & 211068400 & 0.51 & 4.065041 & 207.0 & 3.0 & 5210 & 2.713 & 0.083\\
PELS 072 & 211149600 & 0.678 & 0.314465 & 227.0 & 20.0 & 4634 & 2.081 & 0.104\\
PELS 123 & 211002011 & 0.591 & 7.575758 & 132.0 & 4.0 & 4945 & 1.946 & 0.085\\
PELS 162 & 211067702 & 0.589 & 7.692308 & 44.0 & 4.0 & 4950 & 1.309 & 0.085\\
PELS 189 & 211065162 & 0.567 & 7.575758 & 118.0 & 6.0 & 5011 & 1.948 & 0.083\\
V1289 Tau & 211087059 & 0.509 & 0.6897 & 227.0 & 20.0   & 5215    & 2.850  & 0.095\\     
\hline                                   
\multicolumn{7}{l}{$\dagger$ Primary periods from \citet{Rebull16}}
\end{tabular}
\end{table*}

The sample was initially built from the list of 759 Pleiades members whose rotational period was measured with Kepler K2 by \citet{Rebull16}\footnote{The Kepler K2 field-of-view does not completely cover the Pleiades cluster. A significant number of cluster members are therefore missing from the present study}. As shown there, a very good overall agreement exists between the periods derived from the K2 light curves and those previously reported from ground-based studies \citep[e.g.,][]{Hartman10}. We refrain from considering additional periods available from the latter however, as some may be affected by aliases or harmonics. The periodic sample of \citet{Rebull16} was cross-matched with the compilation of 210 lithium equivalent width measurements published by \citet{Barrado16}. This led to an initial sample of 131 Pleiades members with both a known rotational period and a lithium measurement. 

Based on the derivation of accurate rotational periods from exquisite K2 light curves, \cite{Rebull16} built the rotational period distribution of Pleiades members. This is illustrated in Figure~\ref{protjkfull}, where a large dispersion of rotational periods is seen over the 2MASS color range (J-K$_s$)$\sim$0.5-0.7 mag, corresponding to a spectral type range of about K0-K4 at the age of the Pleiades \citep{Pecaut13}, assuming E(J-K$_s$)=0.02 mag \citep{Rebull16}.  Over this spectral range, 34 periodic Pleiades members had a lithium equivalent width measurement available from the literature while 17 others lacked one. The motivation for this new study was to obtain as complete a sample as possible of early K-type Pleiades dwarfs with both a known rotational period and a measurement of lithium content. As described below, we thus obtained spectra and derived lithium equivalent width estimates for the 17 stars that previously lacked one. 


The final sample analyzed here thus includes 148 Pleiades members with known rotational periods and lithium content, of which 51 are of 
early-K spectral type, where the rotational scatter is quite significant. We further checked the membership probability of the 148 Pleiades candidate members using the estimates published by \citet{Bouy15}, which include both the Tycho and DANCE proper motion surveys. Only one source, Pels 069 is not present in the DANCE proper motion catalog of \citet{Sarro14}, but its short rotational period of 1.76 days supports membership. All but eight of the 147 remaining sources have a membership probability above 0.70, and most indeed close to unity. The eight remaining objects are HII 2984, Pels 173, HII 697, HII 303, HII 885, AK IV-314, and DH 803, which have a zero probability of being members according to the DANCE study, and HII 659 with a probability of 0.44. However, independent evidence suggests that at least some of them are actual members: the TGAS parallax for HII 2984 yields a distance of 137 pc; HII 303, HII 659, HII 697, and HII 885 are strong X-ray sources \citep{Stauffer94, Micela99} and have a radial velocity consistent with membership; AK IV-314 and Pels 173 are also radial velocity members. As described in \citet{Sarro14}, the DANCE proper motion survey is statistical in nature and does allow for an error rate estimate of about 4\%, which would correspond to six objects in our sample. Based on independent evidence of membership for the eight sources discussed above, we elected to retain them in the analysis below.

\section{Observations}

We obtained high-resolution optical spectra of 12 Pleiades early K-type stars at Observatoire de Haute-Provence using the Sophie spectrograph \citep{Perruchot08} over several observing runs in late 2016 and early 2017. The SOPHIE spectrograph was used in the High Efficiency mode, delivering a spectral resolution of 40,000 over the wavelength range 3872-6943\AA. It features an entirely automatic data-reduction pipeline which was used here to get wavelength calibrated one-dimensional spectra. High-resolution spectra were obtained for two additional Pleiades members, Pels 071 and Pels 123, at Keck Observatory using the HIRES spectrograph \citep{Vogt94}. 
We also observed three stars with the fiber-fed echelle spectrograph FIES \citep[see][]{Telting14} mounted at the 2.5 m Nordic Optical Telescope (NOT) in September 2017. FIES covers the spectral range 370-830 nm and we used the low-resolution fiber bundle (R=25,000). The spectra were reduced using the dedicated FIES software,  which  takes  care  of  all  standard  steps, from bias subtraction to wavelength calibration. The Journal of observations is given in Table~\ref{obs}. 
      \begin{figure}
   \centering
    \includegraphics[width=9cm]{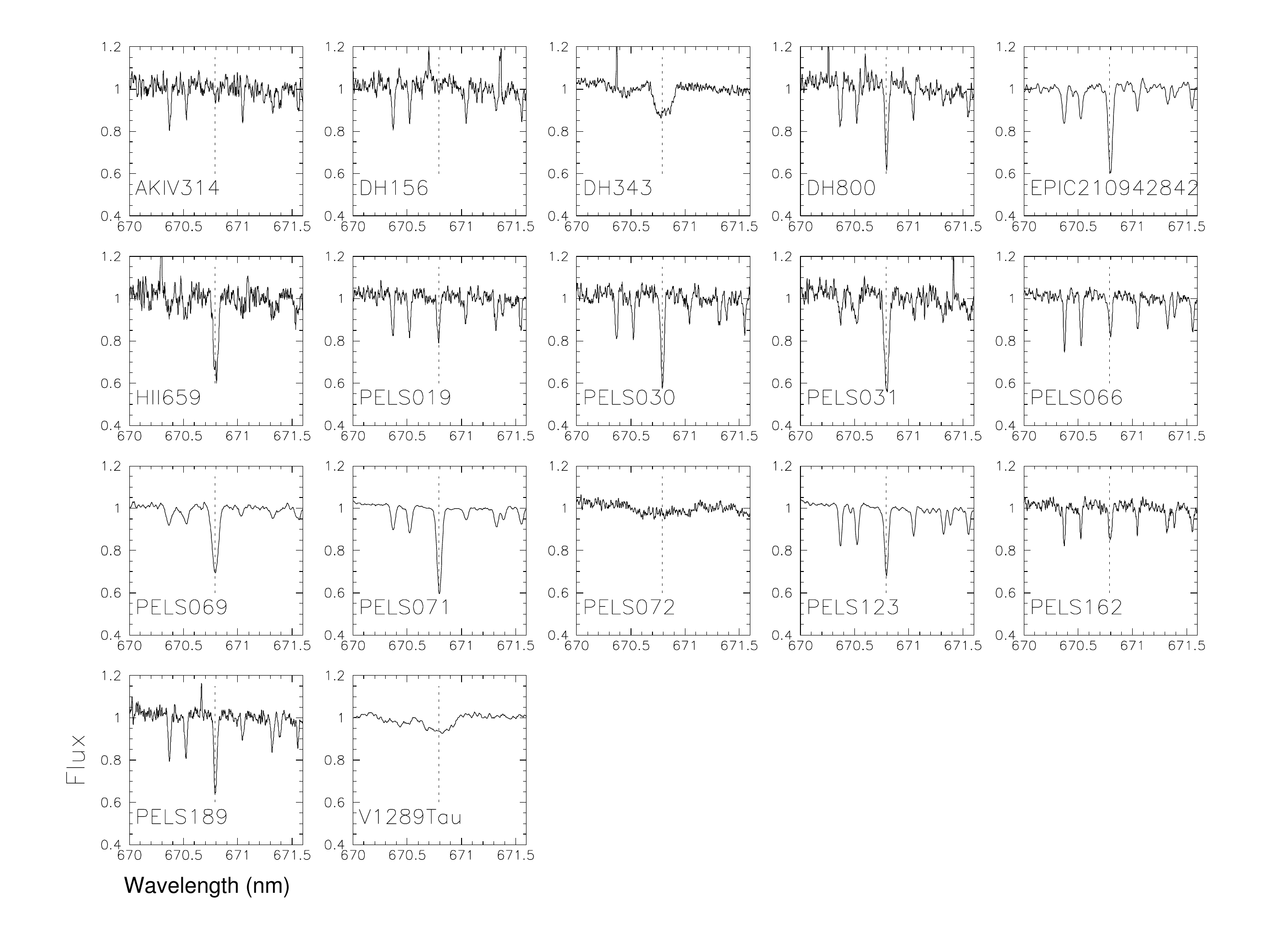}
   \caption{The spectral region around the LiI $\lambda$670.78 nm line is shown for the sample of stars newly observed at OHP, NOT, and Keck. All spectra were flux normalized to unity at the nearby continuum and were boxcar smoothed with a 10-pixel wide filter for clarity (corresponding to 0.01, 0.025, and 0.03 nm for OHP, NOT, and Keck spectra, respectively). The vertical dotted line corresponds to the central wavelength of the LiI 670.78 nm line profile shifted by +5.6 km/s, the average radial velocity of Pleiades cluster members \citep{Galli17}.  } 
              \label{liprof}
    \end{figure}

\section{Results}


     \begin{figure*}
   \centering
    \includegraphics[width=18cm]{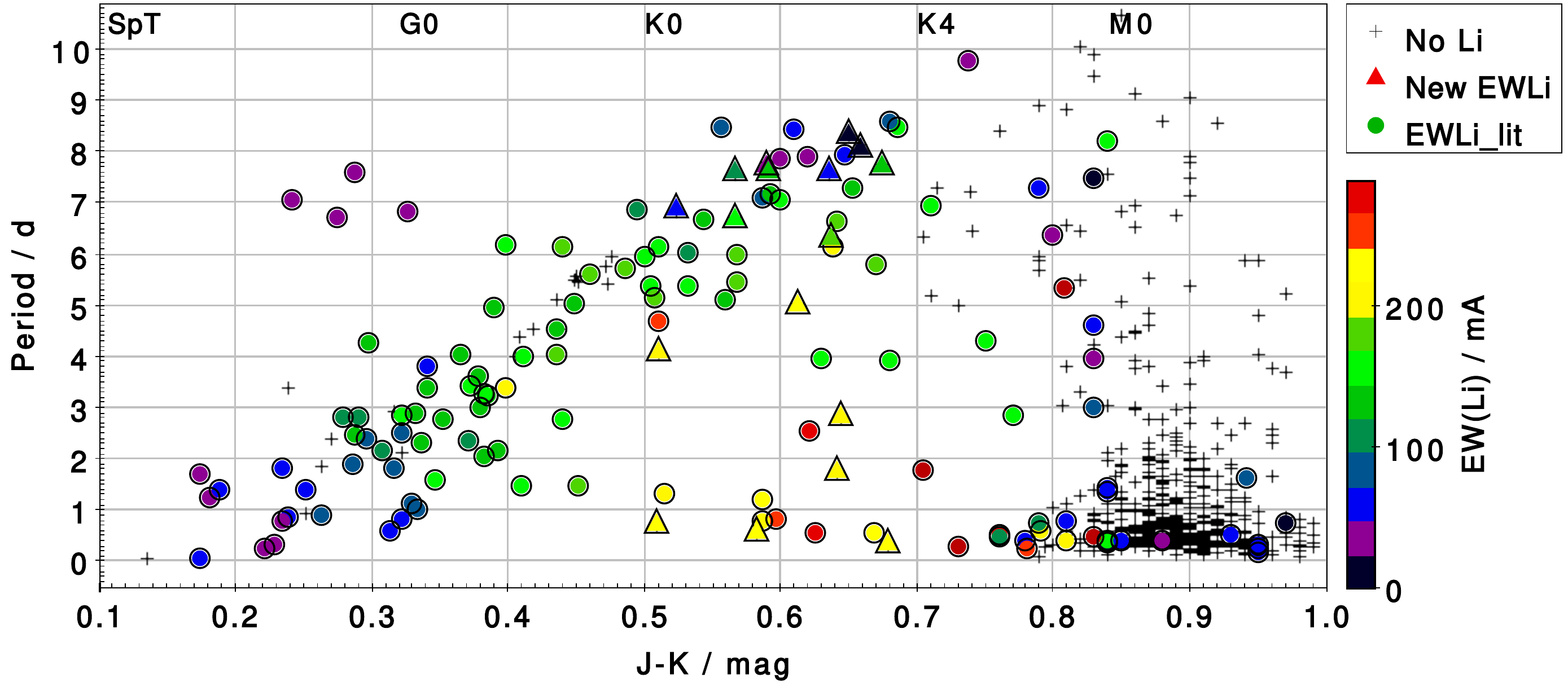}
   \caption{The distribution of K2 rotational periods from \citet{Rebull16} is shown for Pleiades members as a function of (J-K$_s$) color. Crosses are objects without a lithium equivalent width measurement available in the literature, which encompasses mostly late-K- and M-type stars. Circles pertain to LiI 670.8 nm equivalent width measurements gathered from the literature \citep[see][]{Barrado16} while triangles indicate new lithium equivalent width measurements reported in this study. The color of the symbol scales with lithium equivalent width (in m\AA), as shown on the right side of the plot. Approximate spectral types are indicated at the top of the plot. }
              \label{protjkfull}
    \end{figure*}

\subsection{Stellar parameters and lithium abundances}

The equivalent width of the LiI 670.78 nm line was measured on the spectra shown in Figure~\ref{liprof}. Three of us independently performed the measurements, using either direct profile integration or Gaussian fitting.  In most cases, the three independent measurements agreed within the rms error, which amounts to about 5-10 m\AA\ for slow and moderate rotators, and to about 15-20 m\AA\ for fast rotators. We therefore adopted the mean of the three measurements for each object as well as the quadratic mean of their rms error. Before converting lithium equivalent widths to abundances, we corrected the measured LiI 670.78 nm equivalent width from the FeI 607.74 nm blend, adopting the (B-V)-dependent calibration proposed by \citet{Soderblom93}. Over the color range investigated here, the FeI blend correction ranges from 12.8 to 13.8 m\AA. We subtracted a constant value of 13 m\AA\ from the raw LiI equivalent width measurements to correct for this blend. The blend-corrected LiI line equivalent widths are listed in Table~\ref{linew}. For three fast rotators in our sample, DH 343, PELS 072, and V1289 Tau, the LiI equivalent width might be further affected by neighboring metallic line blends (see Fig.\ref{liprof}). To estimate the contribution of these blends in each fast rotator we considered a slow rotator of similar (J-K) colors as a reference spectrum. As a first guess, we renormalized the depth of the narrow LiI line profile of the reference spectrum, so as to recover the raw line equivalent width measured for the fast rotators. We then convolved the reference spectrum by the rotational broadening function of \citet{Gray76} until we found a good fit to the fast rotator spectrum\footnote{We thus used \vsini=60, 80, and 160 km/s for DH 343, V1289 Tau, and PELS072, respectively.}. We finally measured the equivalent width of the LiI line on the non-broadened and on the broadened reference spectra. The difference provides an estimate of the contribution of the blends from neighboring lines to the LiI equivalent width in fast rotators. In the three cases, we found the difference in the measured LiI equivalent widths to be less than 15 m\AA, which is of the order of the measurement uncertainty (cf. Table~\ref{linew}).  We therefore neglected this additional correction for fast rotators.


We used the 2MASS (J-K$_s$) color of the sources, dereddened with E(J-K$_s$)=0.02 mag, to estimate stellar effective temperatures. Using these dereddened (J-K$_s$) colors, we interpolated effective temperatures for 5-30 Myr stars from Table 6 in \citet{Pecaut13}\footnote{The (J-K$_s$) vs. T$_{\rm eff}$ relationship of \citet{Pecaut13} is valid up to (J-K$_s$)=0.91 corresponding to T$_{\rm eff}$=3360 K. For 7 stars in our sample with larger (J-K$_s$) values, up to 1.05 mag, we arbitrarily set T$_{\rm eff}$=3000K.}. Taking into account color uncertainties, stellar variability, and the steep slope of the color-\teff\ relationship, we estimate an average error of 200~K on the \teff\ thus derived\footnote{Using the \citet{Pecaut13} main sequence color-\teff\ relationship instead of the 5-30 Myr one would yield colder \teff\ by 60$\pm$30 K for the stars listed in Table~\ref{linew}. This would translate into lower lithium abundances by an amount of -0.08$\pm$0.04 dex on average.}. 

The effective temperature and associated error were then used to convert lithium equivalent widths to abundances, defined as A(Li)=12+$\log$(Li/H), using the tabulation provided by \citet{Soderblom93} down to an effective temperature of 4000~K.  Errors on the abundances were estimated by bootstrapping using the uncertainties on EW(Li) and T$_{\rm eff}$. The results for the subsample of K-type stars with newly derived lithium equivalent width are listed in Table~\ref{linew}. The results for the whole sample are listed in Table 2 (electronically available from CDS Strasbourg; see a sample in Table~\ref{param}). 

We emphasize that lithium abundances derived from line equivalent widths are subject to significant  systematic uncertainties depending on the choice of the stellar parameters, in particular surface gravity and microturbulent velocity \citep[see, e.g.,][]{Basri91}. Hence, a quantitative comparison between measured lithium abundances and the predictions of stellar evolution models is valid only if the former have been derived using the same atmospheric model as the latter, which is usually not the case. Fortunately, this systematics should not strongly impact the {\it relative} values of lithium abundances at a given effective temperature, and this allows us to investigate the relationship between lithium abundances and stellar rotation at a given mass among the low-mass members of the Pleiades cluster. 

\subsection{The lithium-rotation connection} 

Figure~\ref{protjkfull} shows the distribution of rotational periods and lithium content as a function of the (J-K$_s$) color index for FGK stars in the Pleiades cluster. The new determinations of lithium equivalent widths reported here clearly support the earlier finding by \citet{Soderblom93}, based on a smaller sample and on projected rotational velocities: a clear relationship exists between lithium content and rotation among early K-type stars of the cluster. This is particularly remarkable over the (J-K$_s$) color range from 0.5 to 0.7 mag (corresponding to SpT$\sim$K0-K4, and M$_\star$$\sim$0.75-0.90 \msun\ according to the 125 Myr isochrone of the \citet{Baraffe15} evolutionary models), where the rotational spread is important, and fast rotators exhibit systematically larger lithium equivalent widths than their more slowly rotating counterparts at the same effective temperature. On the blue side, the lithium-rotation connection can be seen down to (J-K$_s$)$\sim$0.50 mag, while it does not appear to be present at bluer colors, in spite of a residual rotational scatter noticeable down into the 0.35-0.45 color range (SpT$\sim$G0-G5, M$_\star$$\sim$1.0-1.2 \msun). On the red side, the relationship may be present as far as (J-K$_s$)$\sim$0.75-0.80 mag (SpT$\sim$K4-K5, M$_\star$$\sim$0.65 \msun), although the lack of lithium measurements for most lower-mass stars prevents us from putting a firm boundary on the red edge of the lithium-rotation connection in the cluster.  

In the color region where the lithium-rotation connection is the most conspicuous, one source appears to depart from the general trend that associates lithium richness with fast rotation and lithium deficiency to low spin rates. HII 263 is located close to the blue edge of the domain, with (J-K$_s$)=0.51 mag and \prot=4.67 days. Despite its relatively moderate rotation rate, it has an EW(Li) amounting to 252 m\AA; it is surrounded in Fig.~\ref{protjkfull} by cluster members of similar rotation rates and spectral type that exhibit significantly lower lithium equivalent widths, in the range from 135 to 190 m\AA. Yet, HII 263's lithium content was independently measured by \citet{Soderblom93} and by \citet{King10}, who consistently report a large EW(Li) of 290 and 252 m\AA, respectively. \citet{Jeffries99} and \citet{Ford02}, however, measured a lower EW(Li) value of 226$\pm$5 m\AA\ and 186$\pm$2 m\AA, respectively. The 2MASS (J-K$_s$) color is consistent with (V-K$_s$)$_0$$\sim$2.08 mag reported by \citet{Rebull16}, both color indices indicating a spectral type about G8-G9 (\teff$\sim$5200 K). \citet{Raboud98} found HII 263 to be a spectroscopic binary with a long-term radial velocity drift. It is however unclear how the binary nature of the source would prevent lithium depletion in a  non-synchronized system\footnote{\citet{Rebull16} report two periods for this source, P$_1$=4.68d and P$_2$=5.11d, which might correspond to the rotational periods of the individual components of the system.} nor how it could induce lithium equivalent width variability.
 Our sample contains a fraction of multiple systems. We present in Appendix C a Figure similar to Fig.~\ref{protjkfull} where known binary systems are identified. We conclude that multiple systems and apparently single Pleiades low-mass members appear to follow the same lithium-rotation relationship.


%
     \begin{figure}
   \centering
    \includegraphics[width=9cm]{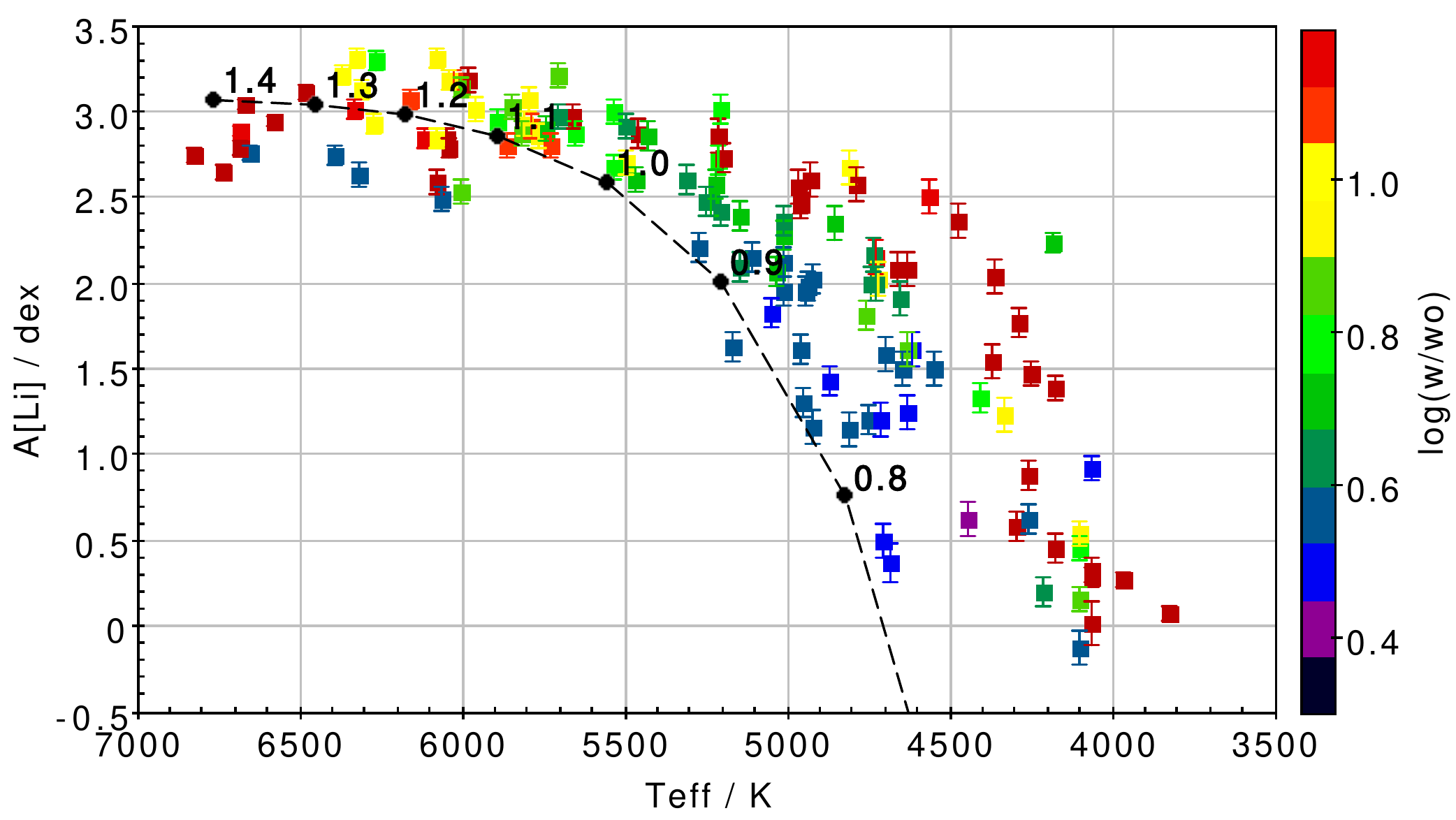}
   \caption{Lithium abundance as a function of stellar effective temperature. The color code reflects the stellar angular velocity normalized to the Sun's on a log scale. The dashed line shows the expected lithium depletion locus at 125 Myr as a function of stellar mass, labeled in units of \msun, as predicted by \citet{Baraffe15} models. } 
              \label{ali}
    \end{figure}

Figure~\ref{ali} shows the derived lithium abundances as a function of effective temperature and angular velocity. While F and early/mid G-type stars at an age of 125 Myr appear to have essentially primordial lithium abundance even well after their arrival on the ZAMS, lower-mass stars have already experienced significant lithium depletion during PMS evolution, at a rate that strongly increases towards the lowest-mass stars. The observed mass-dependent behavior is predicted by PMS evolutionary models \citep[e.g.,][]{Baraffe15}, although an additional mechanism might be required to decrease the amount of lithium depletion predicted by these models at subsolar masses (see Fig.~\ref{ali}). A large scatter of lithium abundance at fixed \teff\ is seen over the range from 4000 to 5300 K and over most of this range, from about 4400 to 5300 K, the lithium abundance appears to be strongly correlated with the stellar spin rate. This further confirms the existence of a tight relationship between lithium content and rotation rate among subsolar-mass stars at an age of 125 Myr over a restricted \teff\ range.

%
 
\section{Discussion}

Most previous studies of the lithium-rotation connection in the Pleiades cluster used \vsini, the projected rotational velocity, as a measure of stellar spin rate. This quantity involves a geometric projection factor which limits its applicability, in particular for low \vsini\ which may correspond to truly slow rotators or moderate/fast rotators seen at a low inclination. Prior to the recent studies of \citet{Gondoin14}, \citet{Barrado16}, and \citet{Somers17}, only \citet{King00} had previously used preliminary estimates of rotational periods derived from the photometric modulation of the stellar luminosity by surface spots. Even though these were much fewer and not nearly as accurate as the Kepler K2 periods used here, \citet{King00} claimed that a few Pleiades K dwarfs with long rotational periods, that is, true slow rotators, were exhibiting Li-excess. The three sources: HII 263, 320, and 1124 are included in the present study. HII 263 was discussed above and we suspect that its spectroscopic binary nature is responsible for its highly variable lithium equivalent width. We argued that the lowest measured value provides the best estimate of the lithium content for the binary components; it is consistent with the expected lithium abundance given the star's rotational period. HII 320 is quoted by \citet{King00} as having a rotational period of 4.58d (although this object does not appear in their table), while the K2 period derived by \citet{Rebull16} is 1.19d, with a secondary period detected at 4.8d. Assuming the shortest period corresponds to the stellar rotational period, the high lithium abundance of this source is fully consistent with fast rotation. Finally, HII 1124,  with a period of 6.12d and (J-K$_s$)=0.64, a lithium equivalent width of 217 m\AA\ and a lithium abundance of 2.2 dex, does not appear to strongly depart from the lithium-rotation relationship seen in Figs.~\ref{protjkfull} and \ref{ali}, respectively. Indeed, Fig.~\ref{protjkfull} shows that all the stars located on the slowly rotating, upper envelope of the period distribution have a lithium equivalent width consistently less than 200 m\AA, while all those with a period of less than 3 days have a lithium equivalent width larger than 200 m\AA\ over the (J-K$_s$)=0.50-0.70 color range. In the intermediate period range, from about 4 to 6 days, the one-to-one correlation between lithium excess and fast rotation may break down. This is seen for instance in a group of  sources over the (J-K$_s$) range 0.60-0.70 mag and rotational periods from 4 to 6 days, where stars with longer periods have a higher lithium content than those with shorter periods. These stars are in the process of migrating from fast to slow rotation, as they are being efficiently braked by strong stellar winds on the zero-age main sequence \citep[e.g.,][]{Reville16}. Whether this transient phase of rapid rotational evolution may trigger additional mixing processes and/or temporarily modify the mixing rate, thus affecting lithium abundances in a specific way, remains to be investigated. Interestingly, \citet{Lanzafame17} report evidence for radius inflation for Pleiades stars precisely over this range of mass and rotation rates. 

Several studies have questioned whether the observed dispersion in lithium equivalent widths among cool stars in young open clusters reflects an intrinsic scatter in lithium abundances or whether it is only apparent, being due to either non-LTE effects in the atmospheres of these active young stars \citep[e.g.,][]{Stuik97} or the consequence of a varying spot coverage from star to star impacting on the LiI 670.8 nm line depth \citep[e.g.,][]{Barrado01}. Because strong magnetic activity is usually associated with fast rotation, the prime causal link between the three quantities, rotation, activity, and lithium strength remains ambiguous \citep[e.g.,][]{Barrado16}. Arguments have been presented in favor of an actual lithium abundance dispersion \citep[e.g.,][]{Soderblom93, King10}, while other studies have warned against any definite conclusion \citep[e.g.,][]{Jeffries99, King00}. We offer here two additional, albeit indirect, arguments in favor of an intrinsic scatter in lithium abundances among K dwarfs in the Pleiades cluster. Firstly, magnetic field measurements of early K-type stars of the age of the Pleiades cluster or close to it have been reported by \citet{Folsom16}. No relationship is found between the large-scale magnetic field strength or topology and stellar rotation rate over a range of rotational periods from 0.4 to 6 days. Indeed, K dwarfs approaching or reaching the zero-age main sequence exhibit an order of magnitude spread in their mean large-scale magnetic field strength that does not appear to correlate with the stellar spin rate. It thus appears unlikely that the lithium-rotation connection could be a mere reflection of an underlying relationship between magnetic field and rotation rate in these young cool dwarfs. We note, however, that a clear relationship exists between X-ray emission and rotation among Pleiades cool dwarfs \citep[e.g.,][]{Stauffer94, Queloz98}. 

\begin{figure}[t]
   \centering
    \includegraphics[width=9cm]{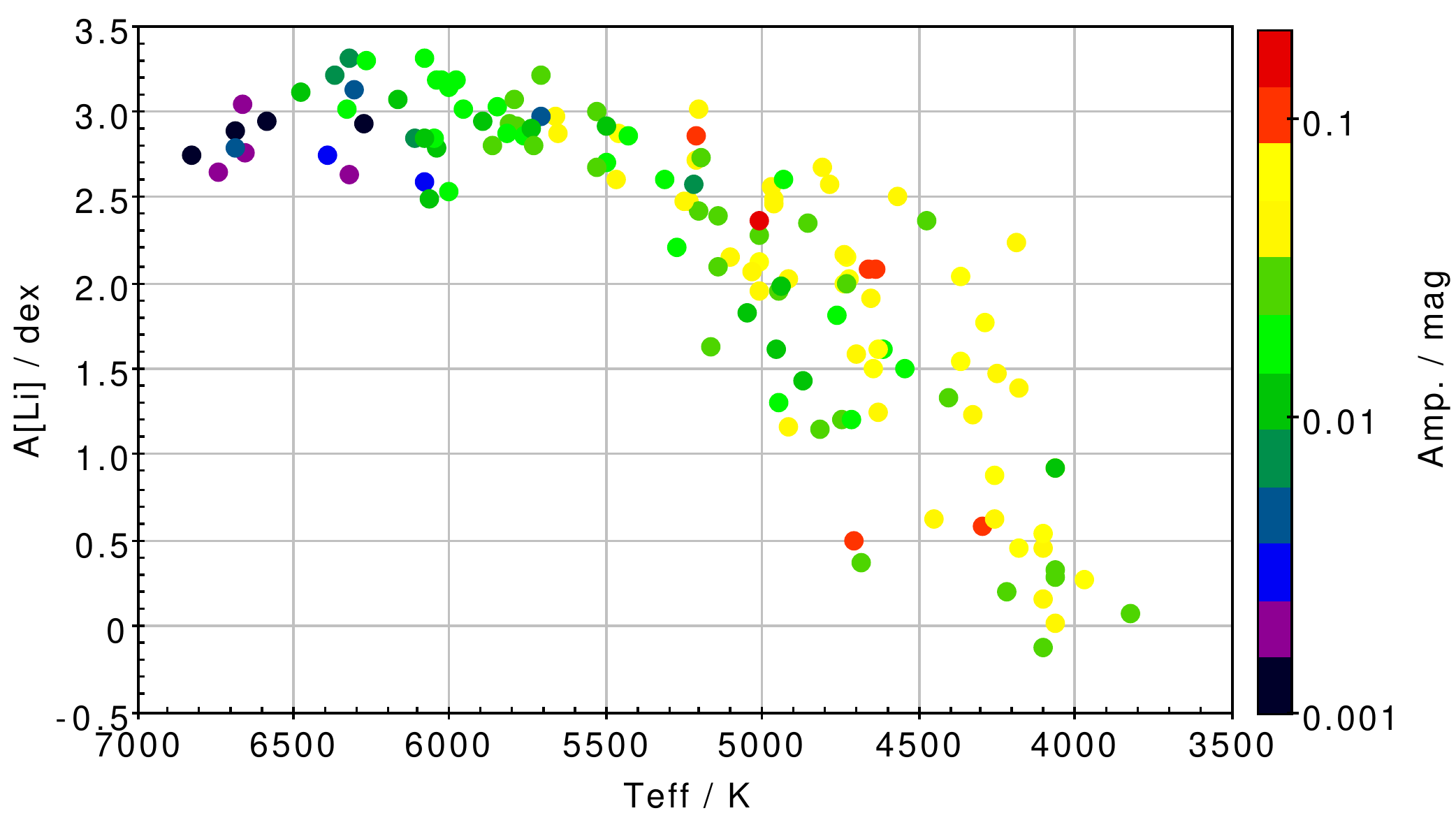}
   \caption{As in Fig.~\ref{ali} except that the color-code now reflects the photometric amplitudes of the K2 light curves, whose log scale is given on the right side of the plot in magnitudes. } 
              \label{amp}
    \end{figure}

\begin{figure}[t]
   \centering
    \includegraphics[width=9cm]{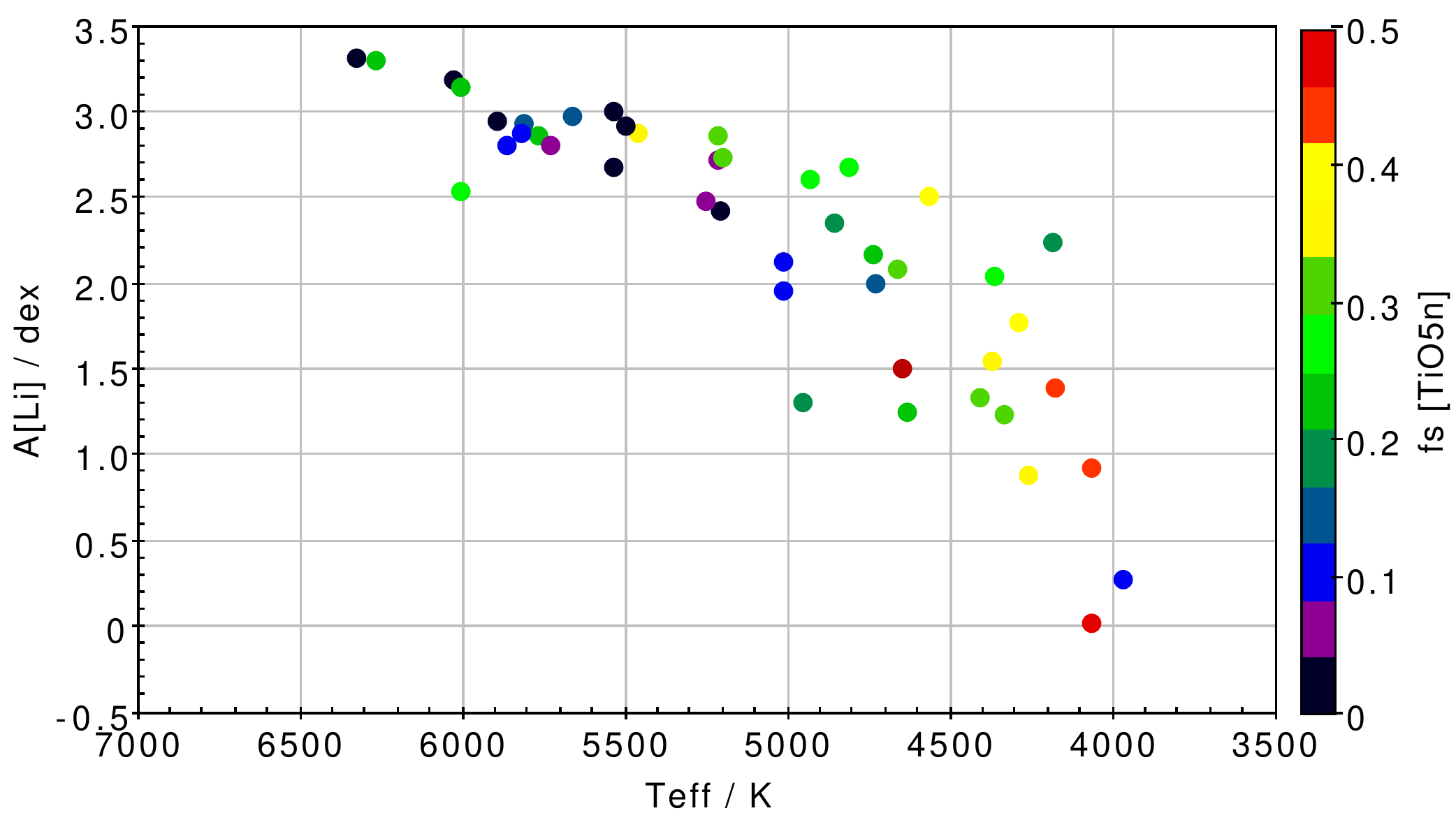}
   \caption{As in Fig.~\ref{ali} except that the color-code now reflects the spot filling factor at the stellar surface, spectroscopically derived from the depth of TiO molecular bands by \citet{Fang16}. The color-scale is given on the right side of the plot. }
              \label{spot}
    \end{figure}

Secondly, we used the photometric amplitude derived from the K2 light curves \citep{Rebull16} as a proxy to surface spot coverage. Figure~\ref{amp} shows lithium abundances as a function of effective temperature and photometric amplitudes. Over the \teff\ range where the lithium-rotation connection is best seen, that is, from about 4400 to 5300 K (see Fig.~\ref{ali}), there seems to be a trend for a correlation between photometric amplitude and lithium abundance, in the sense that stars with the lowest amplitudes of variability seem to concentrate towards the lower envelope of the lithium distribution. However, one also notices a number of lithium-poor stars displaying moderate to large photometric amplitudes over the same \teff\ range. Indeed, the one-to-one relationship clearly seen between lithium abundance and rotation rate in Fig.~\ref{ali} does not seem to have a counterpart in Fig.~\ref{amp}.  Admittedly, photometric amplitudes merely provide a lower limit to spot coverage, as they suffer from projection effects and are only sensitive to the non-axisymmetric part of the spot distribution over the stellar surface. \citet{Fang16} recently reported {\it spectroscopically} measured spot coverage for Pleiades late-type stars, based on the depth of TiO molecular bands, and derived  spot fractional areas covering up to 40\% of the stellar surface for K-type members. Figure~\ref{spot} shows lithium abundances plotted against effective temperature for stars whose areal spot coverage was reported by \citet{Fang16}. Over the \teff\ range from 4500 to 5500~K, no correlation is seen between spot coverage and lithium abundances in this limited sample. Altogether, these results suggest that a differential spot coverage between fast and slow rotators is unlikely to account for the full extent of the lithium dispersion observed among Pleiades K dwarfs, although it might partly contribute to it.



%


Finally, as previously noticed by \citet{Messina16}, the dispersion in lithium equivalent width over the mass range $\sim$0.7-0.9 \msun\ appears to increase by a factor of about two from 5 Myr to 20 Myr (i.e., from NGC 2264 to $\beta$ Pictoris), and again by the same factor between 20 Myr and 125 Myr (i.e., from $\beta$ Pictoris to Pleiades). During most of this pre-main sequence evolutionary phase, low-mass stars tend to remain close to or at a saturated magnetic activity level \citep[e.g.,][]{Pizzolato03,  Wright11}. In contrast, their rotation rate changes quite drastically \citep[e.g.,][]{Gallet13, Gallet15}. Indeed, the growth of the lithium spread during the pre-main sequence mimics the concomitant widening of the rotational distributions up to the zero-age main sequence. In contrast, the strength of stellar magnetic fields steadily declines from the T Tauri phase to the zero-age main sequence \citep[][2017]{Vidotto14, Folsom16}. It is therefore tempting to relate the development of the lithium {\it spread} during the pre-main sequence primarily to the rapid evolution of stellar rotation rates, while enhanced magnetic activity presumably helps to account for inefficient pre-main sequence lithium burning \citep[e.g.,][]{Somers14, Jeffries17}.

\section{Conclusion} 
Fifty years after the first report of an empirical relationship between lithium content and stellar rotation in solar-type stars \citep{Conti68}, the lithium-rotation connection has now been widely documented for low-mass stars. We have gone one step further here in refining the relationship originally reported for low-mass Pleiades members. From a complete sample of 0.75-0.90 \msun\ cluster members, corresponding to spectral types $\sim$K0-K4, we have confirmed that there is a unique relationship between lithium content and rotation rate at 125 Myr, in the sense that faster rotators are systematically more lithium rich than slow ones, with no exception in the investigated effective temperature range. While the interpretation of this observational result is still being debated, additional measurements of lithium abundances and stellar rotational periods are warranted, in particular for late pre-main sequence and young main sequence clusters. This will help to distinguish between the various scenarios proposed to date that attempt to account for the physical cause of the lithium-rotation connection in low-mass dwarfs and its evolution over their lifetime. 

\begin{acknowledgements}
We thank the referee, R. Jeffries, for constructive comments, which helped us improve the manuscript, and A. Lanzafame for useful comments. Discussions with Javier Olivares on Pleiades membership estimates are gratefully acknowledged. We thank the Director of Observatoire de Haute-Provence for the generous allocation of discretionary time after the scheduled observing run had been mostly weathered out, and we are grateful to Melissa Hobson and Flavien Kiefer for obtaining part of the observations presented here. We acknowledge the support of the Programme National de Physique Stellaire of CNRS/INSU. This project was also supported by the Agence Nationale pour la Recherche program ANR 2011 Blanc SIMI 5-6 020 01 ``TOUPIES: TOwards Understanding the sPIn Evolution of Stars''. D.B. acknowledges the Spanish grant ESP2015-65712- C5-1-R. J.B and A.B acknowledge support from CONICYT, Programa MEC 80160028. Partly based on observations made at Observatoire de Haute Provence (CNRS), France. The Nordic Optical Telescope is operated by the Nordic Optical Telescope Scientific Association at the Observatorio del Roque de los Muchachos, La Palma, Spain, of the Instituto de Astrofisica de Canarias. The W. M. Keck Observatory is operated as a scientific partnership among the California Institute of Technology, the University of California and the National Aeronautics and Space Administration. The Observatory was made possible by the generous financial support of the W. M. Keck Foundation. This research has made use of the SIMBAD database, operated at CDS, Strasbourg, France. This publication makes use of VOSA, developed under the Spanish Virtual Observatory project supported from the Spanish MICINN through grant AyA2011-24052. 
\end{acknowledgements}

\bibliographystyle{aa} 
\bibliography{pleiades_li} 

\begin{appendix} 

\section{Journal of the spectroscopic observations}

\begin{table}[h]
\caption{Journal of the spectroscopic observations (ordered as Table~\ref{linew}). }             
\label{obs}      
\centering                          
\begin{tabular}{l l l l l l l l l }        
Name & RA(2000) & Dec(2000) & V & Date Obs. & BJD Obs. & Exp. Time & S/N & Tel. \\
     & {\it (hh mm ss)}  & \it{(dd mm ss)} & \it{(mag)} & \it{(yyyy-mm-dd)} & & \it{(sec)} & \\
\hline                        
EPIC 210942842 & 03 48 46.6 & +22 04 13 & -- &  2017-09-24 & 2458020.7284 & 1800 & 45 & NOT\\
AKIV-314 & 03 54 08.9 & +24 20 01 & 12.48 & 2017-02-05 & 2457790.3877 & 1800 & 24 &OHP\\
DH 156 & 03 40 51.3 & +23 35 54 & 12.79 & 2017-02-16 & 2457801.2984 & 3600 & 28 & OHP\\
DH 343 & 03 44 28.1 & +19 11 06 & 11.84 & 2016-12-05 & 2457728.4750 & 1391 & 59 & OHP\\
DH 800 & 03 54 18.8 & +25 29 43 & 12.87 & 2017-02-15 & 2457800.3027 & 3600 & 30 &OHP\\
HII 659 & 03 45 26.0 & +23 25 49 & 12.23 & 2016-11-23 & 2457716.3988 & 2600 & 19 & OHP\\
PELS 019 & 03 40 30.7 & +24 29 14 & 11.72 & 2017-02-03 & 2457788.3476 & 1282 & 31 & OHP\\
PELS 030 & 03 42 59.2 & +22 54 03 & 12.10 & 2017-02-03 & 2457788.3683 & 1800 & 27 & OHP\\
PELS 031 & 03 43 19.0 & +22 26 57 & 11.49 & 2016-11-23 & 2457716.3051 & 2400 & 23 & OHP\\
PELS 066 & 03 47 34.2 & +21 44 49 & 12.32 & 2016-12-06 & 2457728.5244 & 1500 & 50 & OHP\\
PELS 069 & 03 52 53.0 & +21 46 55 & 11.77 & 2017-09-25 & 2458021.6515 & 2400 & 60 & NOT\\
PELS 071 & 03 53 23.7 & +24 03 54 & 11.40 & 2016-10-14 & 2457675.3956 & 240 & 120 & Keck\\
PELS 072 & 03 50 35.7 & +25 25 35 & 12.10 & 2016-12-10 & 2457733.4468 & 1800 & 52 & OHP\\
PELS 123 & 03 33 13.8 & +23 00 23 & 11.98 & 2016-10-14 & 2457675.3669 & 240 & 101 & Keck \\
PELS 162 & 03 57 33.3 & +24 03 11 & 12.11 & 2016-12-11 & 2457734.4167 & 1500 & 39 &OHP\\
PELS 189 & 03 36 30.3 & +24 00 44 & 12.26 &2017-02-05 & 2457790.3650 & 1500 & 33 & OHP\\
V1289 Tau & 03 54 25.2 & +24 21 36 & 11.05 & 2017-09-24 & 2458020.6888 & 1800 & 55 & NOT\\
\hline                                   
\end{tabular}
\end{table}
 
\section{Table 2. Stellar parameters and lithium content}

Only the first five lines of Table 2 are presented here. The full table is available electronically at CDS Strasbourg.

\begin{table}[h]
\caption{Stellar parameters and lithium content for our sample of 148 stars (first 5 lines only). }             
\label{param}      
\centering                          
\begin{tabular}{l l l l l l l l l l l l l l}        
\hline\hline                 
Name & EPIC & RA(2000) & Dec(2000) & V & J-K$_s$\tablefootmark{a} & Period & Amp. &  WLi & rms & T$_{eff}$ & A[Li] & rms & Bin.\\
   &  & {\it (deg)}   & \it{(deg)} & \it{(mag)} &  \it{(mag)}  & {\it(d)} & \it{(mag)} & \it{m\AA} & \it{m\AA} & \it{K} & \it{dex} & \it{dex}\\
\hline                        
PELS 123 & 211002011 & 53.30793 & 23.00647 & 11.98 & 0.591 & 7.575758 & 0.028 &   132 & 4 & 4945 & 1.95 & 0.084 &\\
PELS 124 & 210990525 & 53.88203 & 22.82362 & 9.86 & 0.289 & 2.7953 & 0.005 &   98 & 10 & 6307 & 3.12 & 0.059 & \\
PELS 189 & 211065162 & 54.12626 & 24.01222 & 12.26 & 0.567 & 7.575758 & 0.035 &   118 & 6 & 5011 & 1.95 & 0.083 & \\
PELS 019 & 211095259 & 55.12802 & 24.4873 & 11.72 & 0.524 & 6.849315 & 0.022 &   49 & 5 & 5168 & 1.62 & 0.079 & \\
DH 156 & 211039307 & 55.21359 & 23.59843 & 12.79 & 0.65 & 8.333334 & 0.085 &  $\leq$15 &  -- & 4708 & 0.50 & 0.092 & \\
...\\
\hline                                   
\end{tabular}
\tablefoot{
\tablefoottext{a}{The Pleiades were one of the regions covered by the 2MASS ``6x'' deeper integrations. The 2MASS measurements were harvested first from the main 2MASS catalog, and then from the deeper integration 6x catalog, allowing the 6x catalog to take precedence over the main 2MASS catalog. In most cases, this is advantageous, because the 6x data have higher signal-to-noise ratios; in a few cases (e.g., HII 335), the 6x data were significantly inferior to the original 2MASS catalog in that their J-K$_s$ colors were inconsistent with their colors from other bands or their H-K$_s$ colors. In such cases, we gave precedence to the 2MASS catalog. }
}
\end{table}
 
\newpage
.

\newpage

\section{Multiple systems }

\begin{figure}[h]
   \centering
    \includegraphics[width=8cm, angle=0]{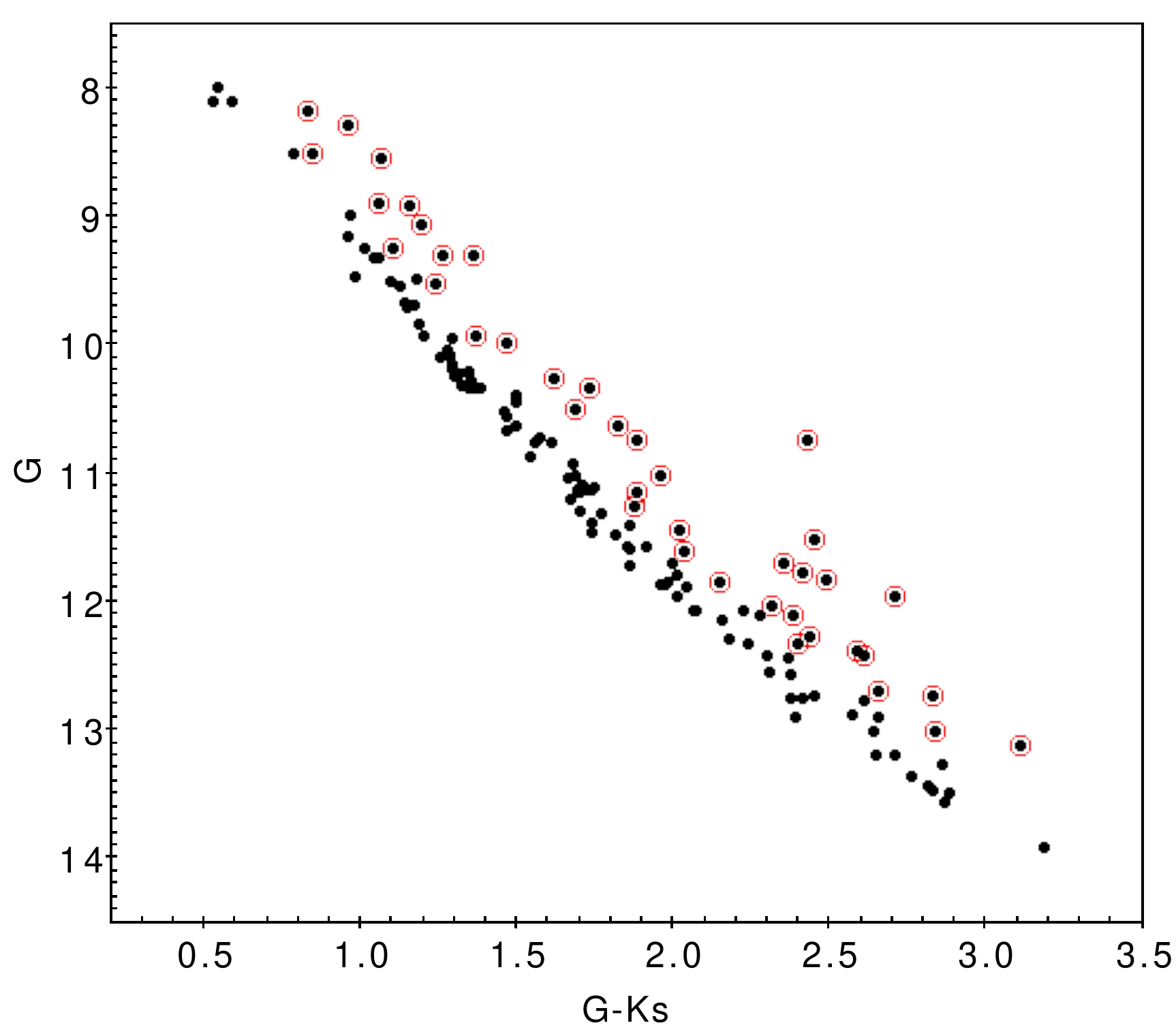}
   \caption{Color magnitude diagram of the sample in the (G-K$_s$, G) plane. Photometric binaries were selected as lying above the single star sequence by more than 3$\sigma$, where $\sigma$ is the rms width of the single star sequence at a given (G-K$_s$) color. Photometric binaries thus identified are show as encircled points in the CMD.  } 
              \label{ggkcmd}
    \end{figure}

We consider here multiple systems included in our sample of Pleiades members with known rotational periods and measured lithium content. The sample was cross-matched with the list of spectroscopic binaries (SBs) reported in \citet{Mermilliod92} and \citet{Raboud98}, and with the sample of spatially resolved visual binaries (VBs) from the adaptive optics study of \citet{Bouvier97}  and speckle imaging from \citet{Hillenbrand17}. In order to further identify photometric binaries (PHBs), we build a ($G$, $G$-K$_s$) color-magnitude diagram for our sample, using Gaia's DR1 for the optical magnitude and 2MASS for the infrared one. The CMD is shown in Fig.~\ref{ggkcmd}. We approximated the single star sequence with two straight lines, on each side of ($G$-K$_s$)=1.7 mag, and considered sources lying above the 3$\sigma$ level from the single star sequence as photometric binary candidates, where we measured $\sigma$=0.1 mag. With thus identified 58 multiple systems in our sample, consisting of 18 SB, 17 VB, and 40 PHB detections (7 SBs and 11 VB's are detected as PHBs), with only one source, HII 2027, identified as a triple SB+VB system. The binary status of the sources is listed in Table~\ref{param}.   

\begin{figure*}[h]
   \centering
    \includegraphics[width=\textwidth, angle=0]{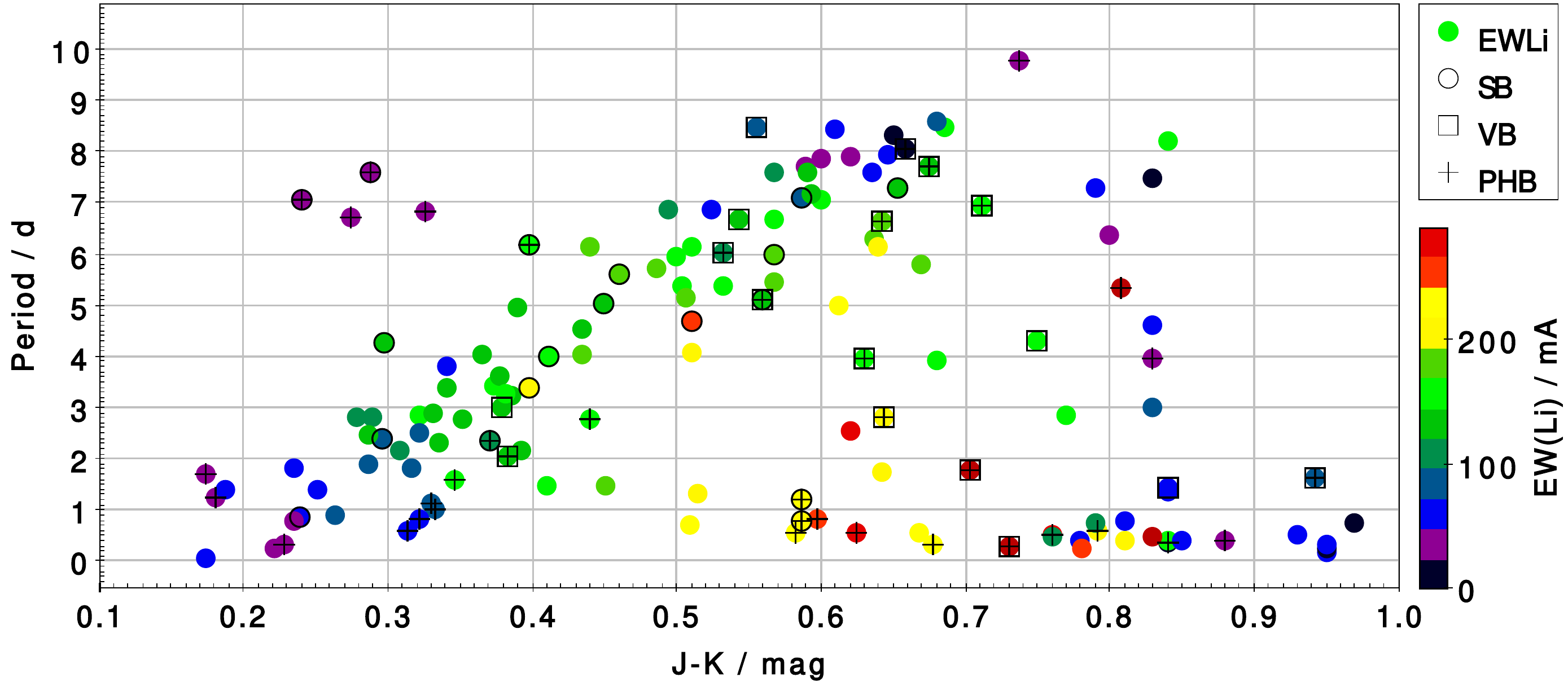}
   \caption{As in Fig.~\ref{protjkfull} but with binary systems identified. Spectroscopic binaries are shown as empty circles, visual binaries as empty squares, and photometric binaries as crosses. } 
              \label{protjkbin}
    \end{figure*}

\begin{figure}[h]
   \centering
    \includegraphics[width=9cm, angle=0]{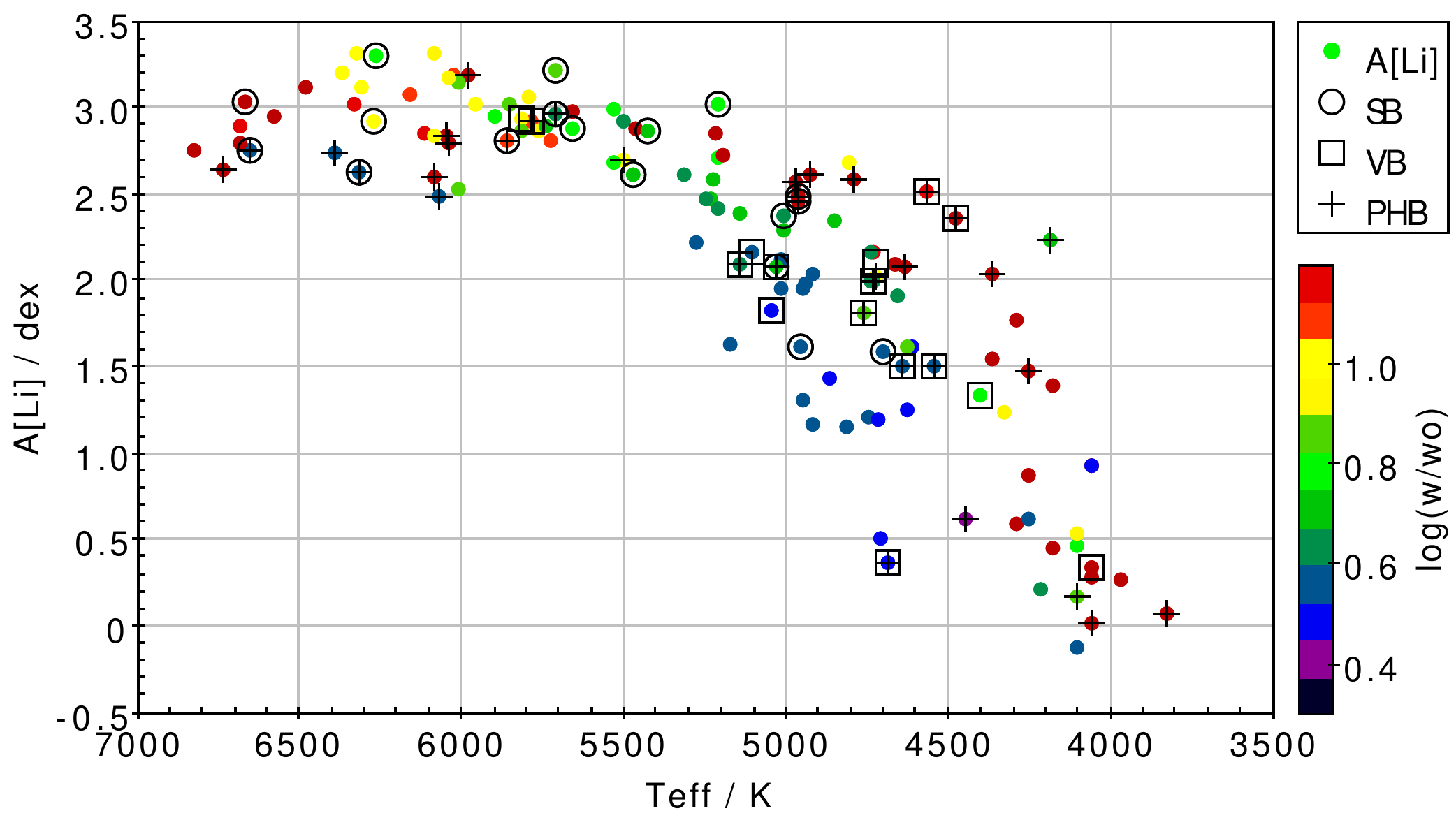}
   \caption{ As in Fig.~\ref{ali} but with binary systems identified. Spectroscopic binaries are shown as empty circles, visual binaries as empty squares, and photometric binaries as crosses. } 
              \label{alijkbin}
    \end{figure}

Figure~\ref{protjkbin} is the same as Fig.~\ref{protjkfull} where the various types of binary systems are identified. Spectroscopic binaries  mostly concentrate along the slow rotator sequence, which might partly result from a detection bias. They all have a lithium equivalent width consistent with that of their non-SB neighbors in the lithium-rotation diagram. Visual, often subarcsecond, binaries, are more evenly distributed, with about half of them on the slow rotator sequence and the other half at shorter rotational periods. They also clearly follow the general trend of a lower lithium content at slower rotation rates. The photometric binaries are the most common in our sample. Of the 40 PHBs identified here, 17 were previously reported as SBs or VBs. Interestingly, those that remain unresolved appear to concentrate at short rotation periods in Fig.~\ref{protjkbin}, at least over the range  (J-K$_s$)$\sim$0.5-0.8 mag. The majority of these unresolved photometric binaries probably have system components with intermediate separations of about 5-20 AU, that is, orbital periods too long to be detected as SBs, while still too tight to be seen as subarcsecond VBs. This suggests that the companion might have disrupted the circumprimary disk early on \citep[e.g.,][]{Cieza09, Messina17}, thus leading to rapid rotation on the zero-age main sequence \citep[e.g.,][]{Gallet13}. We note that among M dwarfs in the Pleiades,
there is a strong correlation between rotation and binarity - such that the
members of dM binaries are more rapidly rotating than dM singles 
as shown in \citet{Stauffer16}. In any case, these systems also follow the general trend of higher lithium abundances at fast spin rates, like their non-binary neighbors in the lithium-rotation diagram. We therefore conclude that low-mass single stars and binary systems alike follow the same lithium-rotation connection in the Pleiades cluster.  For the sake of completeness, we also provide Figure~\ref{alijkbin}, which is the same as Fig.~\ref{ali} with binary systems identified. 

\end{appendix}

\end{document}